\documentclass[fleqn,twoside]{article}
\usepackage{espcrc2}

\usepackage{graphicx}
\usepackage[figuresright]{rotating}

\newcommand{\AmS}{{\protect\the\textfont2
  A\kern-.1667em\lower.5ex\hbox{M}\kern-.125emS}}

\title{A Nearly Model-Independent Characterization of Dark Energy Properties
as a Function of Redshift}

\author{Ruth A. Daly\address{Department of Physics, 
	Pennsylvania State University, Berks Campus, Reading, PA 19610}%
      	\thanks{This work was supported in part by the U. S. National
	Science Foundation (NSF) under grant AST-0507465.}, and 
	S. G. Djorgovski\address{Division of 
	Physics, Mathematics, and Astronomy, 
	California Institute of Technology, MS 105-24, Pasadena, CA 91125}%
	\thanks{This work was supported in part by the NSF under 
	grant AST-0407448 and by the Ajax Foundation.
	We acknowledge the outstanding work and efforts of many observers who
	obtained the valuable data used in this study.}}

\begin{document}

\begin{abstract}
Understanding the acceleration of the universe and 
its cause  
is one of the key problems in physics and cosmology today,
and is best studied using a variety 
of mutually complementary approaches.
Daly and Djorgovski (2003, 2004) 
proposed a model independent approach to determine 
the expansion and acceleration history of the universe and 
a number
of important physical parameters of the dark energy as
functions of redshift  directly from the data.  Here, we apply  
the method to explicitly determine the first 
and second derivatives of the coordinate distance with respect to redshift,
$y^{\prime}$ and $y^{\prime \prime}$, and combine them to solve
for the  
kinetic and potential energy density 
of the dark energy as functions of redshift, $K(z)$ and $V(z)$.  

A data set of 228 supernova and
20 radio galaxy measurements with redshifts from 
zero to 1.79 is used for this study.  
Values of $y^{\prime}$ and $y^{\prime \prime}$ are combined to
study the dimensionless acceleration rate of the 
universe as a function of redshift, $q(z)$. 
The only assumptions underlying our determination
of $q(z)$ are that the universe is described 
by a Robertson-Walker (RW) 
metric and is spatially flat. 
We find that the universe is accelerating today, and was decelerating in the
recent past.  The transition from acceleration to deceleration occurs
at a redshift of about $z_T = 0.42 \pm {}^{0.08}_{0.06}$.  
Values of $y^{\prime}$ and $y^{\prime \prime}$ are combined to 
determine $K(z)$ and $V(z)$.  These are shown to be consistent with
the values expected in a standard Lambda Cold Dark Matter (LCDM) model.  
\end{abstract}

\maketitle

The acceleration of the universe at the present epoch has been 
studied in the contexts of specific models 
using coordinate distances to type Ia supernovae 
\cite{R98,P99,T03,K03,B04,R04,A05}, and FRII radio galaxies 
\cite{DGW98,GDW00,DG02,PDMR03}, in addition to other techniques. 
These studies indicate that
the universe is expanding at an accelerting rate at the present
epoch. Generally, these studies are done in the context of a specific
cosmological model, such as an open universe with non-relativistic matter,
a cosmological constant, and space curvature (e.g. 
\cite{R98,P99,DGW98,GDW00}), 
a spatially flat universe
with non-relativistic matter and 
dark energy that has an energy density that can evolve with redshift but
which maintains a constant equation of state (e.g. 
\cite{P99,T03,K03,B04,R04,A05,DG02}), or
a spatially flat universe with non-relativistic matter and an 
evolving scalar field (e.g. \cite{PR00,PDMR03}).   
In each of these studies it is implicitly assumed that the universe
is desribed by a RW metric and that General Relativity is the correct
theory of gravity; in addition, a particular functional form for 
the redshift evolution of the energy density of some new component is 
assumed. The data are then used to constrain the parameters that describe
the assumed functional form for the redshift evolution of whatever was
being considered as the driver of the acceleration of the universe.

\begin{figure}
\includegraphics[width=70mm]{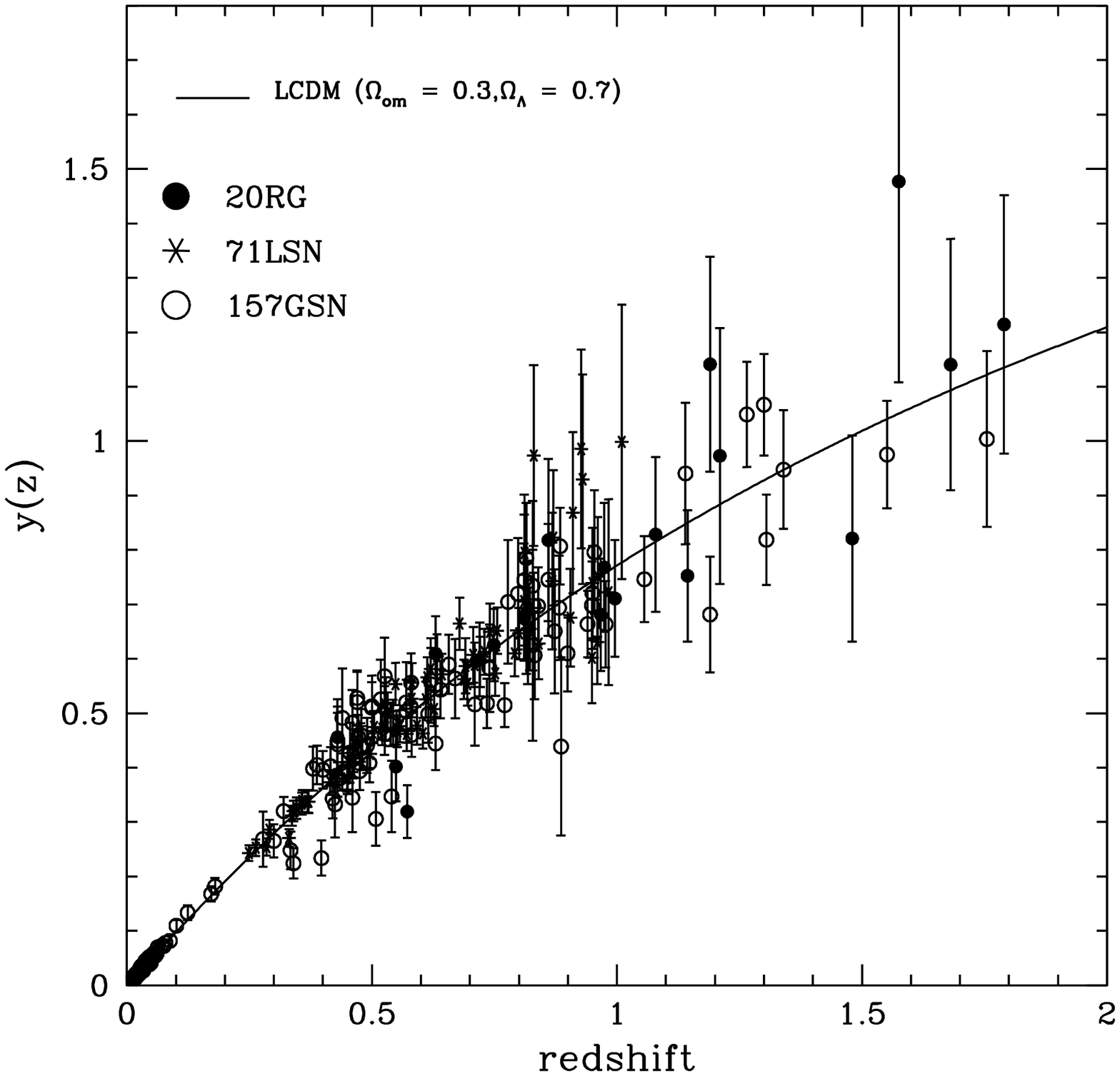}
\caption{\label{Y248} Dimensionless coordinate distance $y(z)$ 
to 71 Legacy and 
157 Gold supernovae, and 20 radio galaxies;  
$y(z)\equiv H_0(a_0r)$, 
$H_0$ is Hubble's constant, and $(a_0r)$ is the coordinate distance
to a source at redshift z. 
It is convenient to work with $y(z)$ because 
it is independent of $H_0$ (i.e. $(a_0r) \propto H_0^{-1}$, 
so $H_0(a_0r)$ is indepenent of $H_0$).  }

\includegraphics[width=70mm]{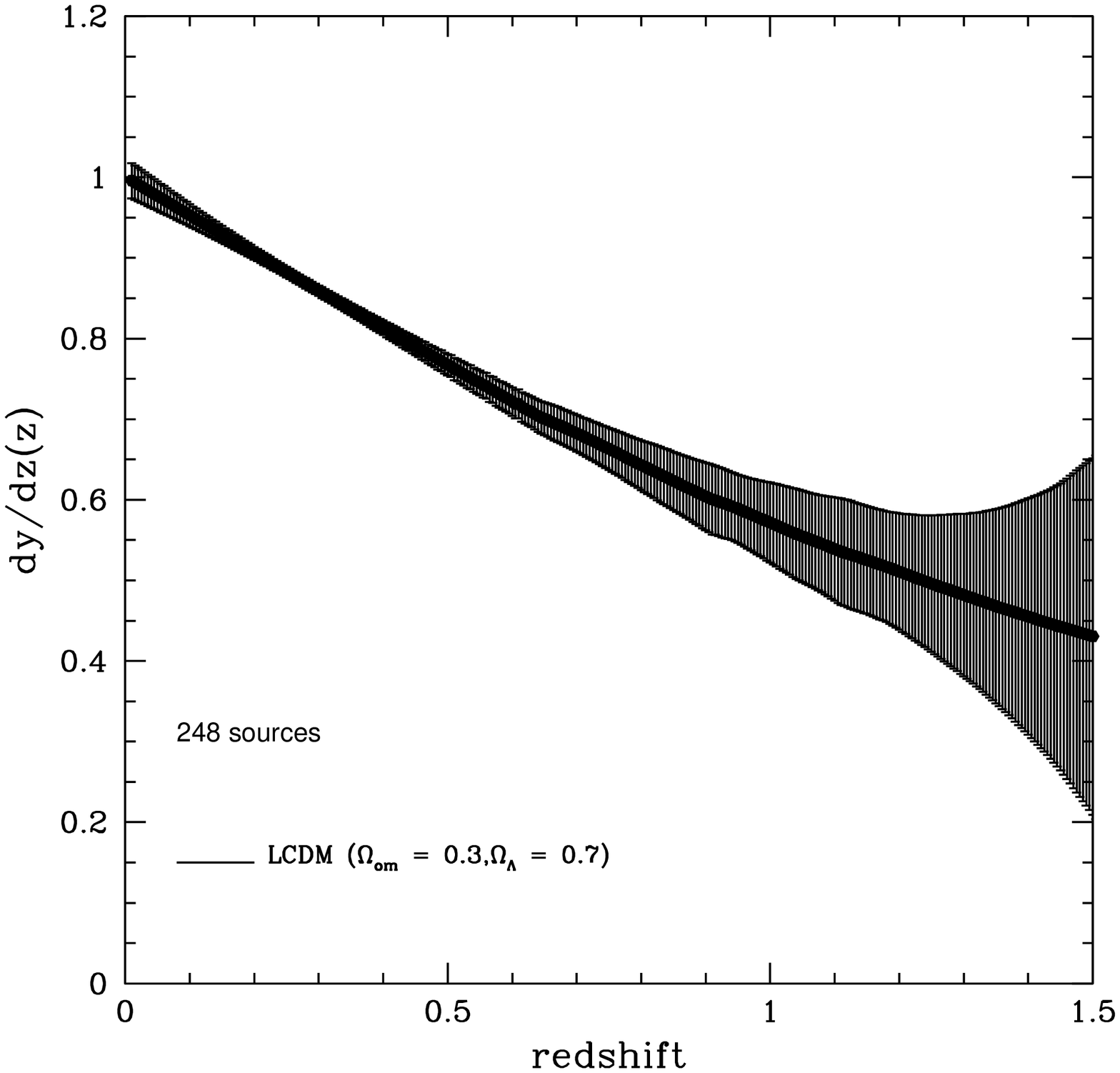} 
\caption{\label{dydz248compare} Results obtained with  
the mock data set of 248 sources described in the text. 
The results are in an excellent agreement
with the input cosmolgoy, with no apparent bias.}
\end{figure}

\begin{figure}
\includegraphics[width=70mm]{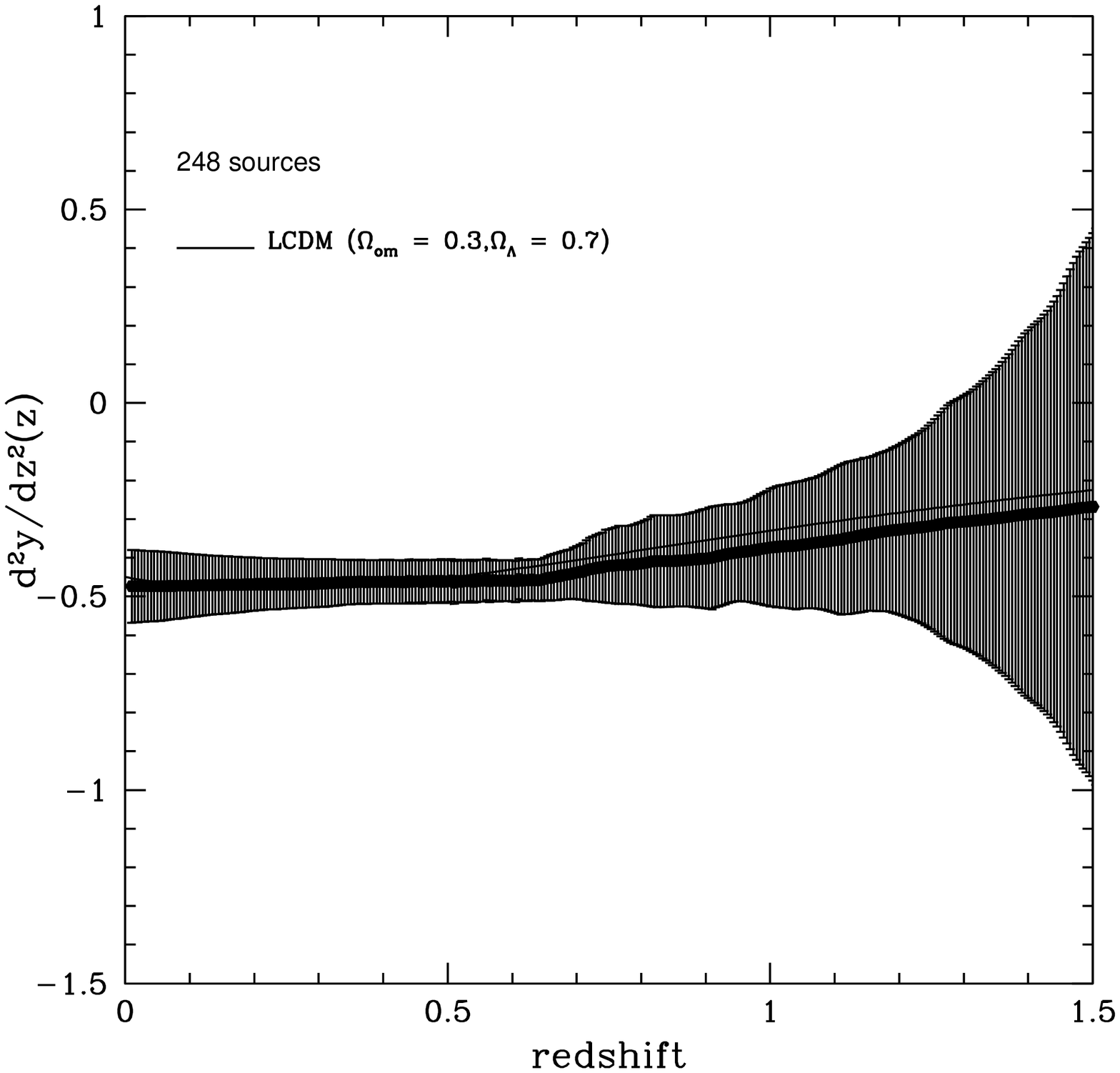}
\caption{\label{d2ydz2compare248} 
As in Fig. \ref{dydz248compare} for $y^{\prime \prime}$.  
The correct assumed cosmology is recovered with a
negligible bias.}

\includegraphics[width=70mm]{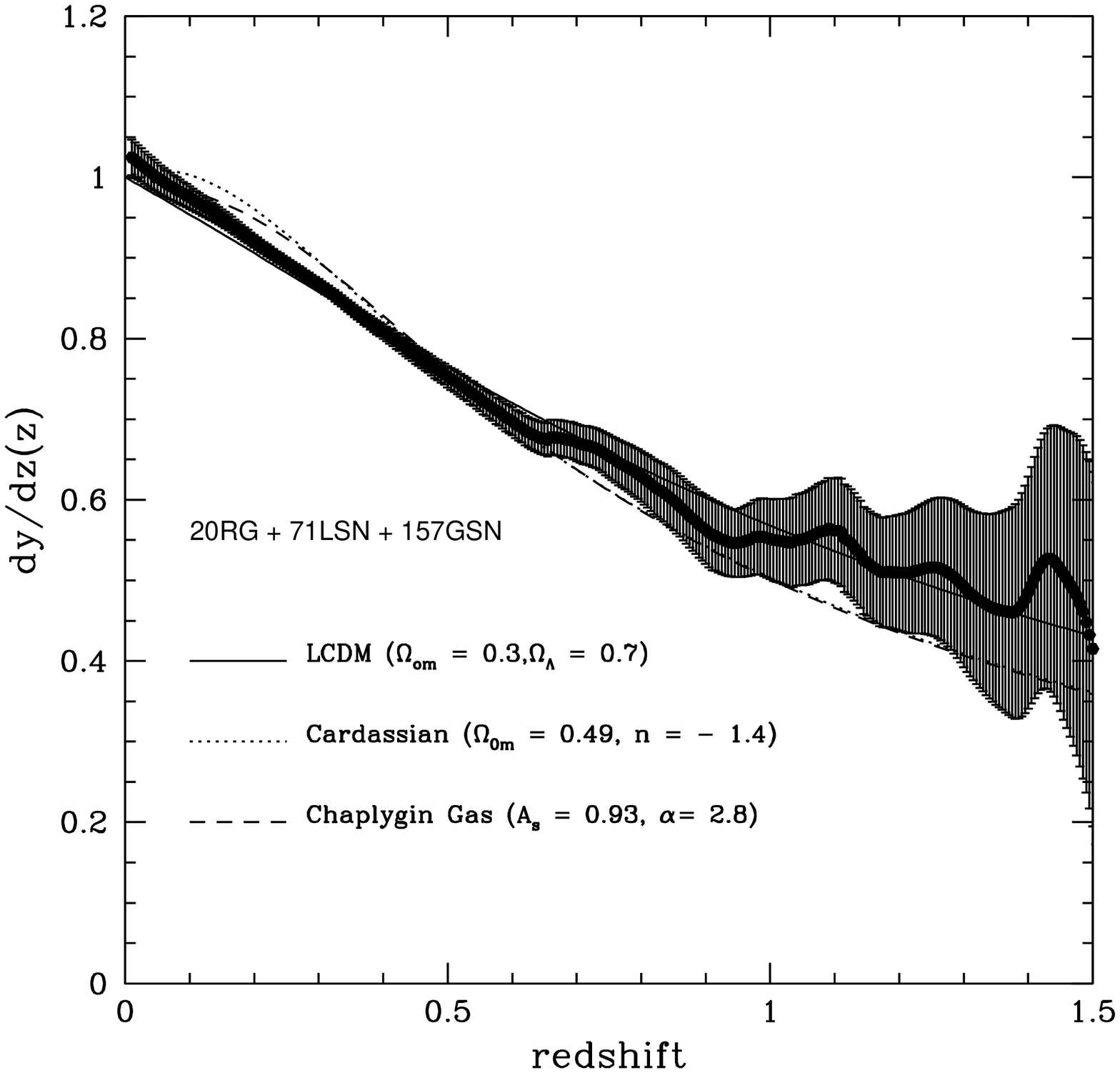} 
\caption{\label{dydz248snrg} The first derivative of the coordinate 
distance
with respect to redshift for the actual data set of 248 sources.  
The zero redshift
value we measure is $y ^\prime _0 = 1.025 \pm 0.022$;
the predicted value in all models is 1.000.  
The values for the standard LCDM model with 
$\Omega_{\Lambda} = 0.7$ and $\Omega_{0m}=0.3$ are shown as the
solid line in this and all subsequent plots.
Best fit Cardassian (dotted line) and Chaplygin Gas 
(dashed line) models are also shown, and
are described in the text.   } 
\end{figure}

\begin{figure}
\includegraphics[width=70mm]{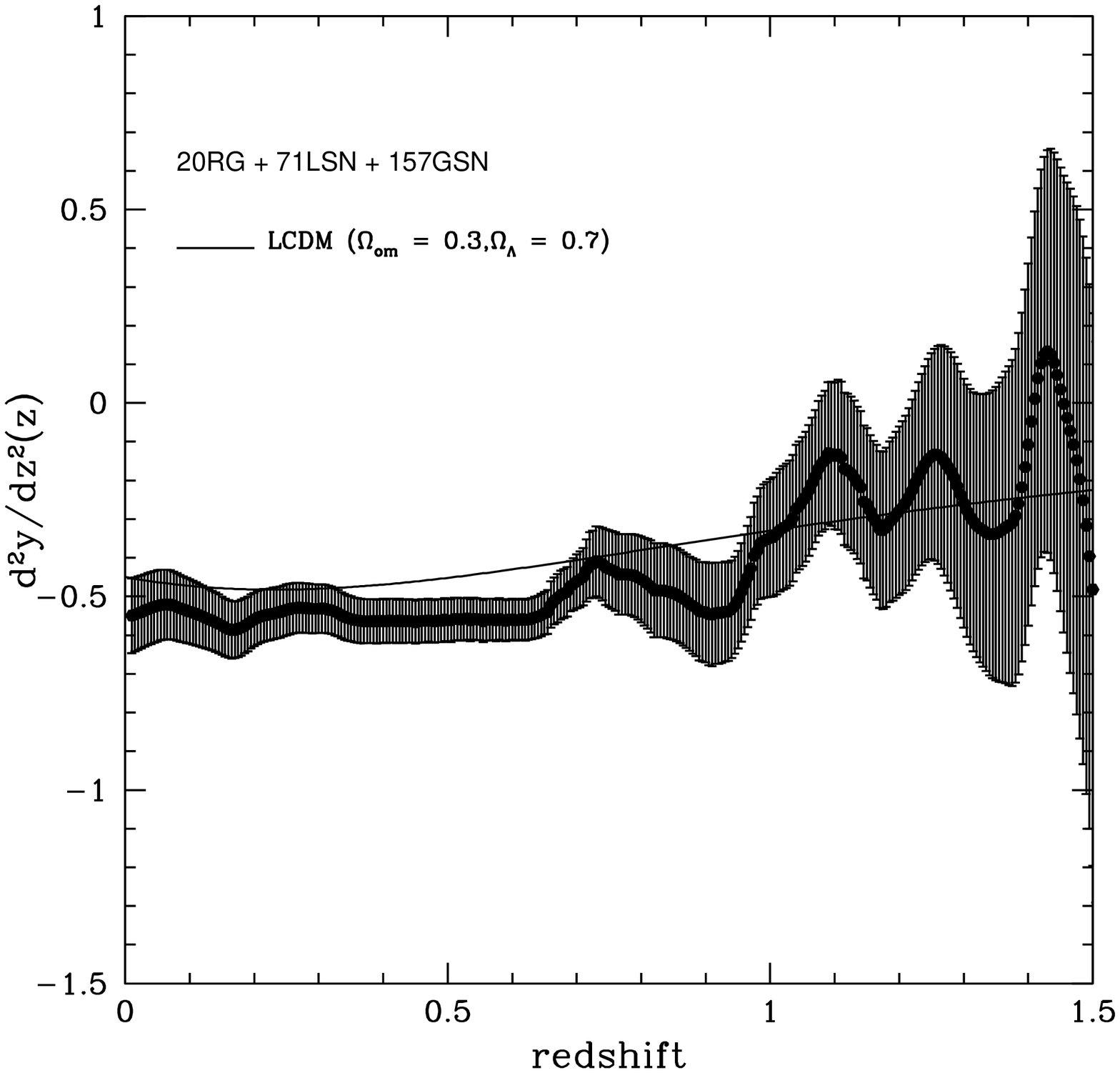}
\caption{\label{d2ydz2snrg248}The second derivative of the 
coordinate distance with 
respect to redshift for the actual data set of 248 sources. 
The measured zero redshift
value is $y ^{\prime \prime} _0 = -0.55 \pm 0.10$; the value predicted
in the LCDM model shown is $-0.45$.}

\includegraphics[width=70mm]{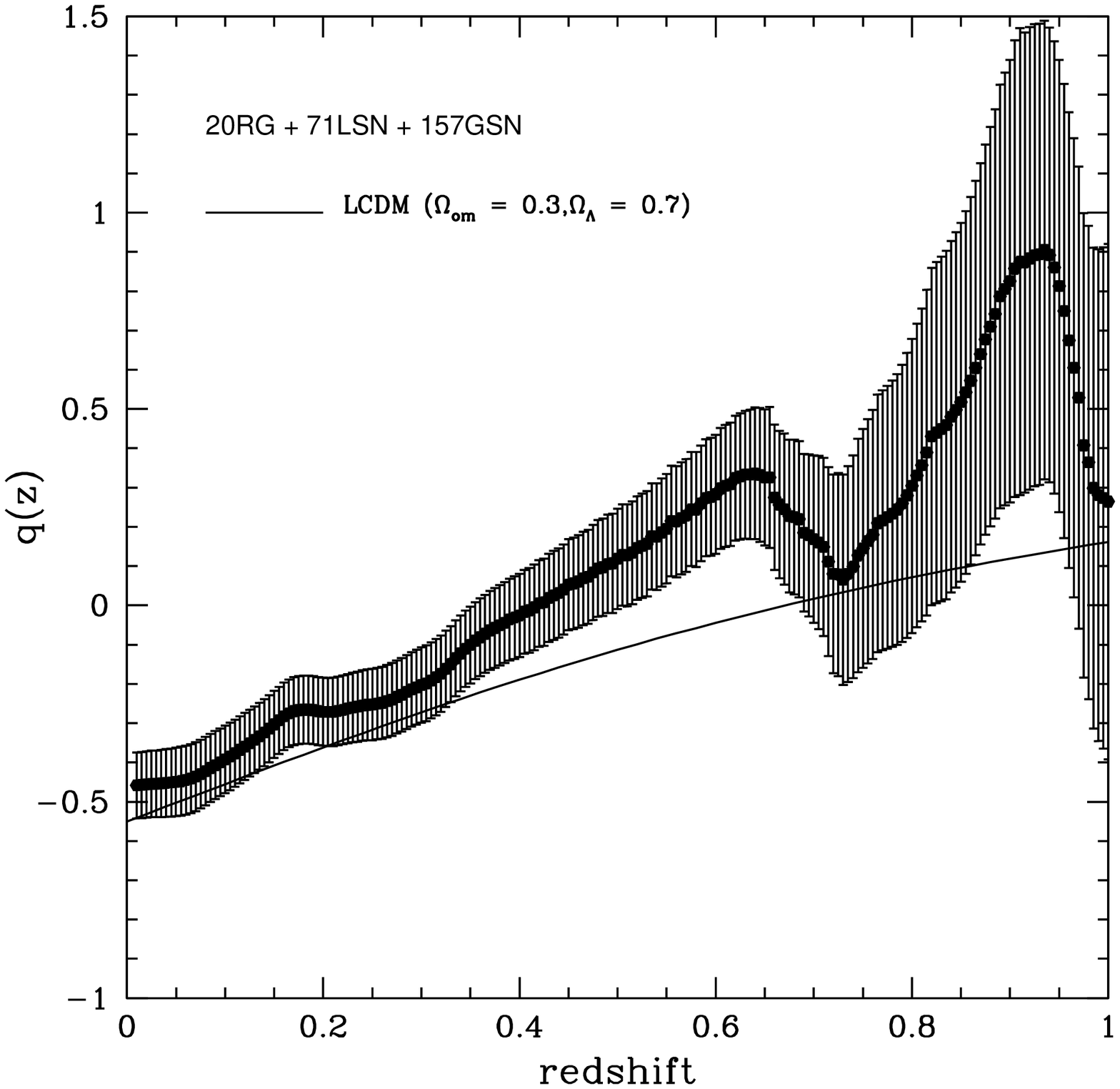} 
\caption{\label{Q248} The deceleration parameter $q(z)$, where
$q(z) = -[1+y^{\prime \prime}(1+z)/y^{\prime}]$ \cite{dd03}.  
The zero redshift
value is $q_0 = -0.46 \pm 0.08$. The predicted value in the 
LCDM model shown is $-0.55$.
Our fits are systematically higher than the LCDM model shown  
by about 1$\sigma$.} 
\end{figure}

A complementary approach was suggested by \cite{dd03,dd04} who showed
that the recent expansion and acceleration history of the universe, and 
some properties of the driver of the accelertion, can be determined directly
from the data after specifying a minimal number of assumptions.  
Assuming only that
the universe is described by a RW metric and is spatially
flat, the data can be used to solve for the dimensionless expansion 
and acceleration rates of the universe as functions of redshift, 
$E(z)$ and $q(z)$, respectively.  This can be done without specifying a 
theory of gravity, or anything else.  The function $q(z)$ thus obtained
is a direct measure of the acceleration/deceleration of the universe at 
different epochs. 
The key ingredients that go into the determination of
$E(z)$ and $q(z)$ are the first and second derivatives of the coordinate
distance with respect to redshift, $dy/dz$ (or $y^\prime$)  and $d^2y/dz^2$
(or $y^{\prime \prime}$), 
which are obtained from the 
coordinate distances to supernovae and radio galaxies at known redshift, 
as described by \cite{dd03,dd04}. Thus, rather than assuming a functional
form for the redshift evolution of the ``dark energy'' and constraining
the model parameters, it is possible to solve for quantities such as $q(z)$ 
directly.  

This direct approach indicates that the universe is accelerating today, and
was decelerating in the recent past. The data used
for the results shown here include 
the sample of 157 ``Gold'' supernovae
\cite{R04}, the sample of 71 supernvae 
from the Supernova Legacy Survey \cite{A05},
and the 20 radio galaxies of \cite{GDW00}, as described in detail 
\cite{DD05}.   
The total sample of 248 sources 
is shown in Fig. \ref{Y248}; there
are no systematic differences seen among the three groups of measurements
in the redshift ranges of their overlaps.

\begin{figure}
\includegraphics[width=70mm]{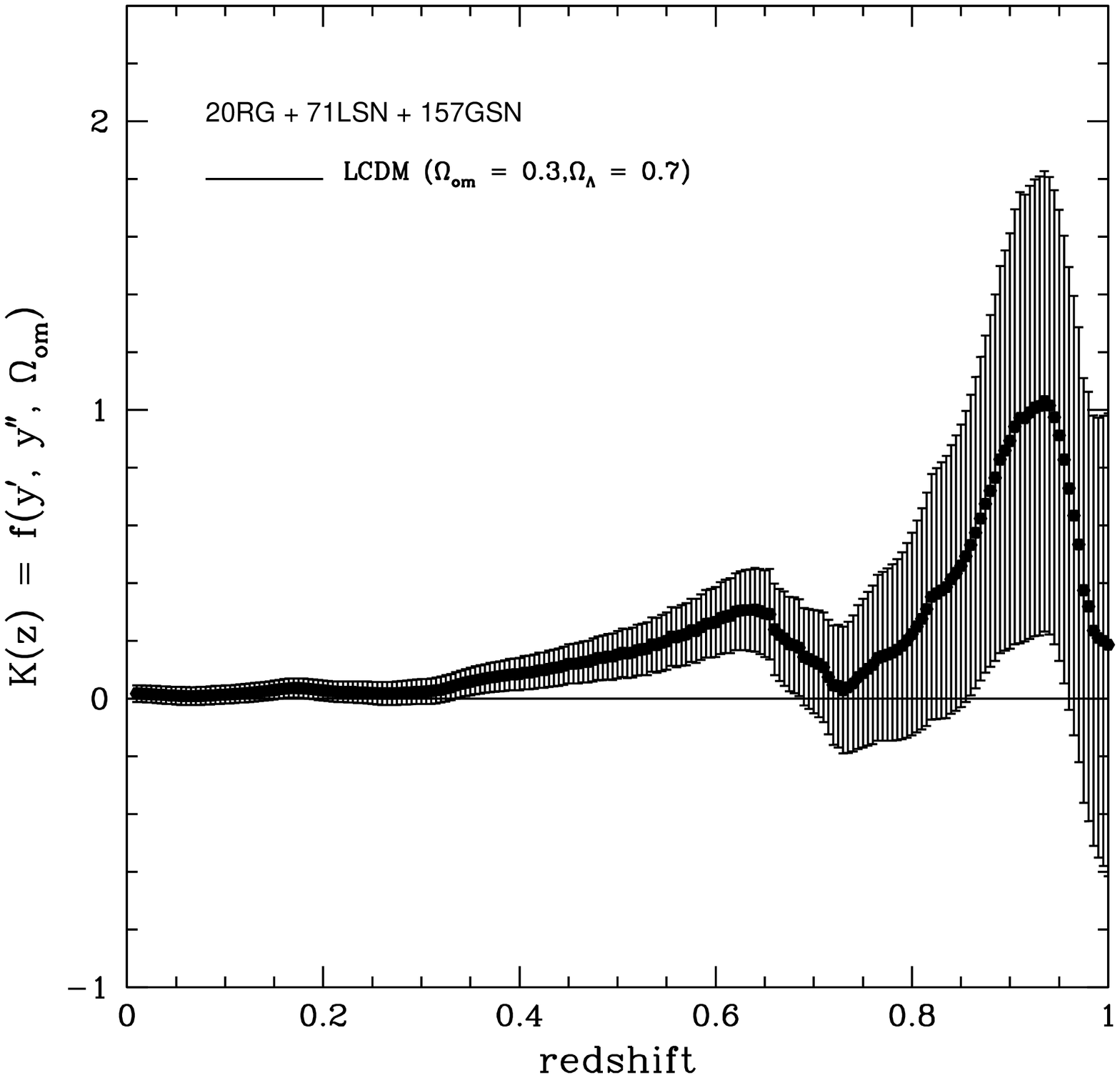} 
\caption{\label{K248} The kinetic energy  of the dark
energy K(z) in units of the critical density today. 
The zero redshift value is $K_0 = 0.02 \pm 0.03$; the expected value for the   
LCDM model shown 0.} 

\includegraphics[width=70mm]{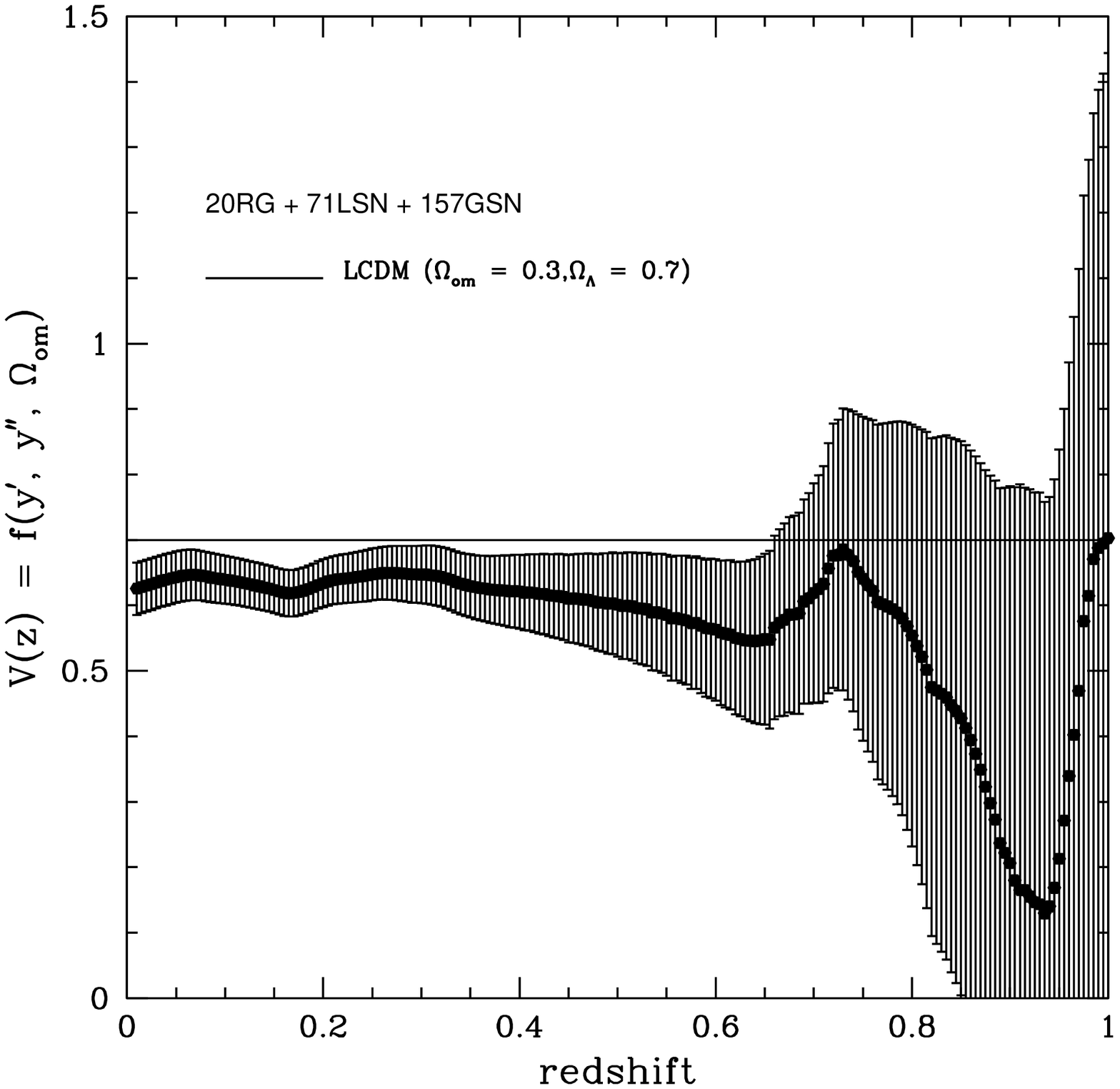} 
\caption{\label{V248} The potential energy of the dark energy 
$V(z)$ in units of the critical density today. The zero
redshift value is $V_0 = 0.62 \pm 0.05$; the expected
value for the LCDM model shown is 0.7.} 
\end{figure}

The first and second derivatives of the coordinate distance with respect
to redshift are obtained using the numerical differentiation method
described by \cite{dd03,dd04}. To test whether the method introduces
a bias in the results, a mock data set of 248 sources with the same  
redshift distribution and 
fractional uncertainty per point as the actual data was  
constructed assuming a LCDM model with $\Omega_{0m}$ = 0.3 and 
$\Omega_{\Lambda}=0.7$, and analyzed.  The results are shown in 
Figs. [\ref{dydz248compare}] and [\ref{d2ydz2compare248}].
We see that no bias has been 
introduced by the numerical differentiation technique.

Values of $y^{\prime}$ and $y^{\prime \prime}$ are shown 
on Figs. [\ref{dydz248snrg}] and [\ref{d2ydz2snrg248}]. 
The ringing seen in these figures is most
likely due to sparse sampling.  
In these plots, and in the ones that
follow, we do not consider these fluctuations at higher redshifts
to be statistically significant, as they are commensurate
with our derived 1-$\sigma$ error bars. 
The results are consistent with 
the LCDM model.  The LCDM model is based on General Relativity 
with non-relativistic matter
$\Omega_{0m}$ and a cosmological constant. This provides an excellent
description of the data.     
Curves showing predictions 
in two modified gravity models in a spatially flat universe
are also shown on Fig. [\ref{dydz248snrg}].  
These curves shown are obtained using the best fit model parameters obtained
by \cite{BBSS05} for the Cardassian model of \cite{FL02}
and the generalized Chaplygin gas model of \cite{BBS02} based
on the model of \cite{KMP01}; this is consistent with 
the results obtained by \cite{LNP05}. Clearly, the LCDM model provides
a better description of the data than do either of the modified
gravity models.  Thus, this large-scale test of General Relativity 
shows that GR provides an
excellent description of the data on these very large length scales of about 
10 billion light years.  

The deceleration parameter $q(z)$ is shown on Fig.[\ref{Q248}].
These results allow
a determination of the redshift at which  
the universe transitions from an accelerating phase
to a decelerating phase; we find this redshift 
to be $z_T = 0.42 \pm {}^{0.08}_{0.06}$, consistent
with the values quoted by \cite{R04} and \cite{dd03,dd04}. 
The upper bound on this transition redshift is uncertain
because of the fluctuations in $q(z)$ which are due to 
sparse sampling at high redshift.

It is well known that $K = 0.5 (\rho+P)$ and $V = 0.5 (\rho-P)$, 
where $\rho$ and $P$ are the
energy density and pressure of the dark energy.  
In \cite{dd04} we show that 
both $\rho$ and $P$ may be written in terms of the first and
second derivatives of the coordinate distance. 
Combining these, we find that 
$({K/\rho_{oc}}) = -{(1+z)} (y^{\prime \prime})(y^\prime)^{-3}/3-0.5
{\Omega_{0m}}(1+z)^3$, and 
$({V/\rho_{oc}}) =(y^\prime)^{-2}[1+
{(1+z)}y^{\prime \prime}(y^\prime)^{-1}/3] - 0.5{\Omega_{0m}} (1+z)^3~$, 
where $\rho_{oc}$ is the critical density at the current epoch.  These
are shown in Figs. [\ref{K248}] and [\ref{V248}].  
In obtaining $K$ and $V$, the assumptions made to 
obtain $P$ and $\rho$ apply: 
the universe
is spatially flat; the kinematics of the universe are accurately
described by general relativity; and two components, the dark
energy and non-relativisitc matter (with $\Omega_{0m} = 0.3$), 
are sufficient to 
account for the kinematics of the universe out to redshift of about 2
(see the discussion in \cite{dd04}). Functional forms for $P(z)$
and $\rho(z)$ for the dark energy are {\it not} assumed, nor is any assumtion
made regarding the equation of state of the dark energy.  
The work presented here on the potential energy, $V(z)$, 
is complementary to the
work of \cite{SRSS00,CTC04,SVJ05,SLP05}.

Thus, our (nearly) model-independent method provides results which are
consistent with those from the more traditional approaches, in a
largely complementary fashion; at the very least, it is a new way
of looking at the data.  As the quality and size of relevant data
sets increase, we can expect even more useful constraints to 
emerge from this approach.

\end{document}